# THE SLS BEAMLINES DATA ACQUISITION AND CONTROL SYSTEM

J.Krempasky, R.Krempaska, D.Vermeulen, D.Maden,
T.Korhonnen, W.Portmann, S.Hunt, R.Abela,
PSI-SLS, Villigen, Switzerland,
M.Muntwiler, Physik Institut, ETHZ, Switzerland

*Abstract*

On December 15th the Swiss Light Source (SLS) produced a stored beam for the first time. This important milestone was achieved in a very tight time schedule. The fact that all major systems are controlled by Epics made this challenge feasible. In the first phase there are four beamlines: two for the surface science community, one for powder and surface diffraction and computed micro-tomography, and the last one for protein crystallography. All of them are equipped with insertion devices, which users want to treat as active sub-systems like a monochromator or experimental station. The beamline control systems are based on the same hardware and software technology as is the machine. This implies extensive use of Personal Computers running Linux RedHat 6.2 and VME systems (PowerPC). The advantage of this choice is a staightforward implementation of the insertion devices into the beamline and experiment framework. Although the experiment Application Program Interfaces differ from beamline to beamline, the standard software technology for linking all sub-systems is based on the Epics toolkit and Cdev. The diagnostic tools provided by this toolkit are being extensively used during the beamline commissioning. Finallly we account on some examples of integrating dedicated 3rd party and commercial non Epics software products for experiment control into the beamline control system. Key elements in this domain are CORBA, Python and Portable Channel Access Server.

## 1 INTRODUCTION

In the first phase SLS (Swiss Light Source) will have four beamlines. Two for hard X-rays (Protein Crystallography and Material Science) and two for the soft VUV region (Surfaces/Interfaces Spectroscopy and Surfaces/Interfaces Microscopy), all equipped with insertion devices (see Fig. 1). Their commissioning started in April 2001 and by the end of November 2001 all of them succeeded to get the beam in their experimental environments. This very tight time schedule clearly indicates, that the EPICS control system toolkit provided a valuable help for the persons involved in the beamline commissioning.

An important aspect of the SLS beamlines is the fact, that they benefit from the same EPICS based control system as the SLS machine. This involves both EPICS hardware and software technology and allowed us a smooth integration of several sub-systems (monochromator, insertion device, beam diagnostics etc ...).

Despite the fact, that the experimental stations may be based on a non EPICS hardware and software technology, there are several possibilities how to include them into an EPICS system. In this context we will briefly describe a CCD camera distributed architecture, and the EPICS implementation of the Scienta-Gammadata electron analyser for the Surfaces/Interfaces Spectroscopy beamline.

## 2 DATA ACQUISITION AND CONTROL

The SLS beamlines support for controls and data acquisition is integrated into the machine control system. It is based on EPICS, a client-server toolkit with CA (Channel Access) servers running on VME processors (PowerPC), and clients running Linux (RedHat 6.2) or NT PCs [1]. The position and functionality of the VME crates follow the beamline topology (typically front-end, monochomator, beamline optics, experimental station 1, 2...) and clearly determine a beamline sub-system (see Fig. 2). The total number of VME crates per beamline is 4-5. Some sub-systems have been (will be) delivered already with EPICS drivers, for example the Jenoptik Plane Grating Monochromator and COPHEE (COmplete PHotoEmission Experiment) experimental station. Their inclusion into the beamline EPICS environment is straightforward, which greatly facilitates and speeds up the beamline commissioning.

### 2.1 Usage of synApps package on the SLS Beamlines

The bulk of custom synchrotron radiation software is taken from synApps package [2]. It includes software support for motors, scalers, optical tables, slits, multidimensional scans, multichannel analyzers and miscellaneous devices like optical encoders, D/A converters, temperature controllers, etc... The support for motors (both stepping and servo motors) provided by the EPICS motor record, fits well to Oregon Micro Systems VME58 family of intelligent motion controls, generally used for motion control on all beamlines. A monochromator can be considered as a bunch of such motors. They can be driven according to special tables or analythic expressions defined by EPICS *calc* records. The control of the Kohzu double-crystal monochromator for the Protein crystallography beamline (also included in synApps) is a complex state notation program which drives the motors to required positions.

Yet another very useful support from synApps is the so called *sscan* EPICS record. It allows for dynamically configuring various types of scans based on EPICS process variables. This feature is of particular interest for beamline commissioning where one can perform a run-time configuration of 1D/2D scans with several positioners, detectors and detector triggers. The scans can be automatically saved from the VME IOC to a file server - an other very useful feature called *saveData* from the synApps package (there is one dedicated file server for each beamline). Simply put, one can perform simple or sophisticated scans without involving coding on some clients. All data are kept in VME crates and saved automatically. The only limitation is the size of the *sscan* arrays, which is 16K.

## 2.2 Usage of SDDS on the SLS Beamlines

The Material Science Beamline at the SLS is currently making extensive use of the SDDS (Self Describing Data Set) [3] utilities and EPICS client-based applications developed at the APS for beamline tuning and data acquisition scans of the powder diffractometer. In particular, the GUI interfaces to the SDDS software, such as quickExperiment [4], have found acceptance by the beamline scientists during the early stages of beam commissioning. At a later date, it is intended that more science oriented software should be introduced. This will also most likely be coupled with a move to the NeXus [5], which is strongly supported by the neutron scattering community at PSI.

## 3 IMPLEMENTING NON EPICS HARDWARE

### 3.1 CCD camera distributed architecture

An experiment based on a CCD camera data acquisition and control is typically based on a very heterogenous hardware and software architecture. In collaboration with ELETTRA, an open system has been developed, where different software components related to CCD data acquisition and evaluation, are integrated into a single GUI by means of a CORBA architecture. A CCD camera CORBA server implements all functionalities of the CCD camera and is running on a Windows NT operating system. The client provides a simple and intuitive graphical user interface giving a user the possibility to control the CCD camera by initializing, setting or getting CCD parameters, or displaying an image. The user interface provides also a utility for simple image processing like adjusting contrast and brightness, create a histogram of an image or provide arithmetic operations on images. It is implemented in Java. Typically an experiment based on CCD is a sequence of simple operations for the implementation of which a scripting language is welcome. The bulk of operations on CCD camera and EPICS control system is implemented in a class called "Operations" (see Fig. 4). By interpreting the script the reference of this class is passed to the script, which can call the class methods. In such a way a physicist can make an image based experiment in an automatic way with understandable simple scripts.

### 3.2 Controlling the Scienta electron analyser

The heart of the experimental apparatus for the Surfaces/Interfaces Spectroscopy beamline is a Scienta-Gammadata electron analyser. The control philosophy for this analyser (as supplied by the company), is a rather monolythic application. On one side it allows to supply additional functionality, but on the other side it didn't fit into a client-server model. This was a serious obstacle not just for implementing experiments, where the analyser is considered to be just one part of the whole experimental setup, but also for synchronising the Scienta scans with the SLS top-up operation mode. Namely, the analyser is supposed to get into a stand-by mode during the top-up injection. Moreover the application was written in Delphi - a clearly non standard programming language for the SLS controls group.

The use of the Portable Channel Access (ActiveXCA) [6] for invoking the analyser operation-specific functions from the Scienta native code, turned out to be the most simple solution. The Pascal units for the EPICS CA were provided by means of the ActiveX automation interface. The additional coding required just the initialization of the EPICS process variables and implementing their event handlers. In this way we planted EPICS process variables into this Delphi application, which allows us now to see the Scienta analyser as a virtual EPICS IOC. Since the planted EPICS process variables have also the monitor capabilities, we get automatically a server-like behavior. The event handlers implemented inside the planted EPICS process variables invoke the same functions the user invoke by clicking on the buttons in the Scienta native code.

## 4 DATA STORAGE AND HANDLING

In view of the different amounts and rates of data a flexible scheme for the acquisition, storage as well as handling of the data has been developed. The different beamlines produce data at rates ranging from several Mbytes/hour to 80 Mbytes/sec. The total amount of accumulated data ranges from Gbytes to several hundred Gbytes per day/experiment.

Each beamline has a dedicated file server of up to 300 Gbyte of available storage, which can be easily extended to more than 1 Tbyte. Since the beamline is controlled from a private network, the file server is equipped with two network cards that are connected to the private SLS-network and the PSI network, respectively (see Fig. 3). All data has to go through this file server that has to leave the SLS.

Currently a tape-library is being set-up as a central medium-term storage facility. It has a storage capacity of 30 Tbytes. Users can, upon request, transfer their data to this medium. Also, they have the possibility to make copies of their data to other mass-storage media like DLT or DAT.

For the large volume data (more than 100 Gbyte) and commercial users, Network Attached Storage (NAS) or hot-swappable disks can be hooked up to the data-acquisition computers.

User access to the computers on the private SLS-net is granted through individual group accounts, which are connected to individual accounts on the PSI-domain through the group-id. The users at the beamline can decide on the level of security for their data.

## 5 CONCLUSIONS

The choice for the EPICS control system toolkit on the SLS beamlines clearly helps us to meet the tight time schedule of their commissioning. The generic purpose solutions provided by EPICS toolkit in principle avoid any coding. Thus the software maintenance across the beamlines are minimal. Yet another very important aspect is the fact, that even beamline operators with no or minimal knowledge about EPICS, can actively participate on their commissioning. The goal of the beamline controls group is to keep this philosophy valid also for the experimental stations. Provided the special instrumentation functionalities live as EPICS process variables, it is possible to implement various experiments in terms of simple scripts instead of coding special purpose applications. The wide variety of available APIs (tcl/Tk, IDL, Octave, perl ...) for EPICS clearly stem for this solution. Since interfacing to EPICS on Win32 is also possible with ActiveXCA, there is also a big opportunity for the Delphi, VisualBasic and VisualC++ developpers providing also a "bridge" to COM based applications.

## 6 ACKNOWLEDGEMENTS

Many thanks go to Tim Mooney from APS for many valuable discussions and suggestions concerning the synApps package. The authors also wishes to acknowledge all contributions from the partners in the ELETTRA beamline control team.

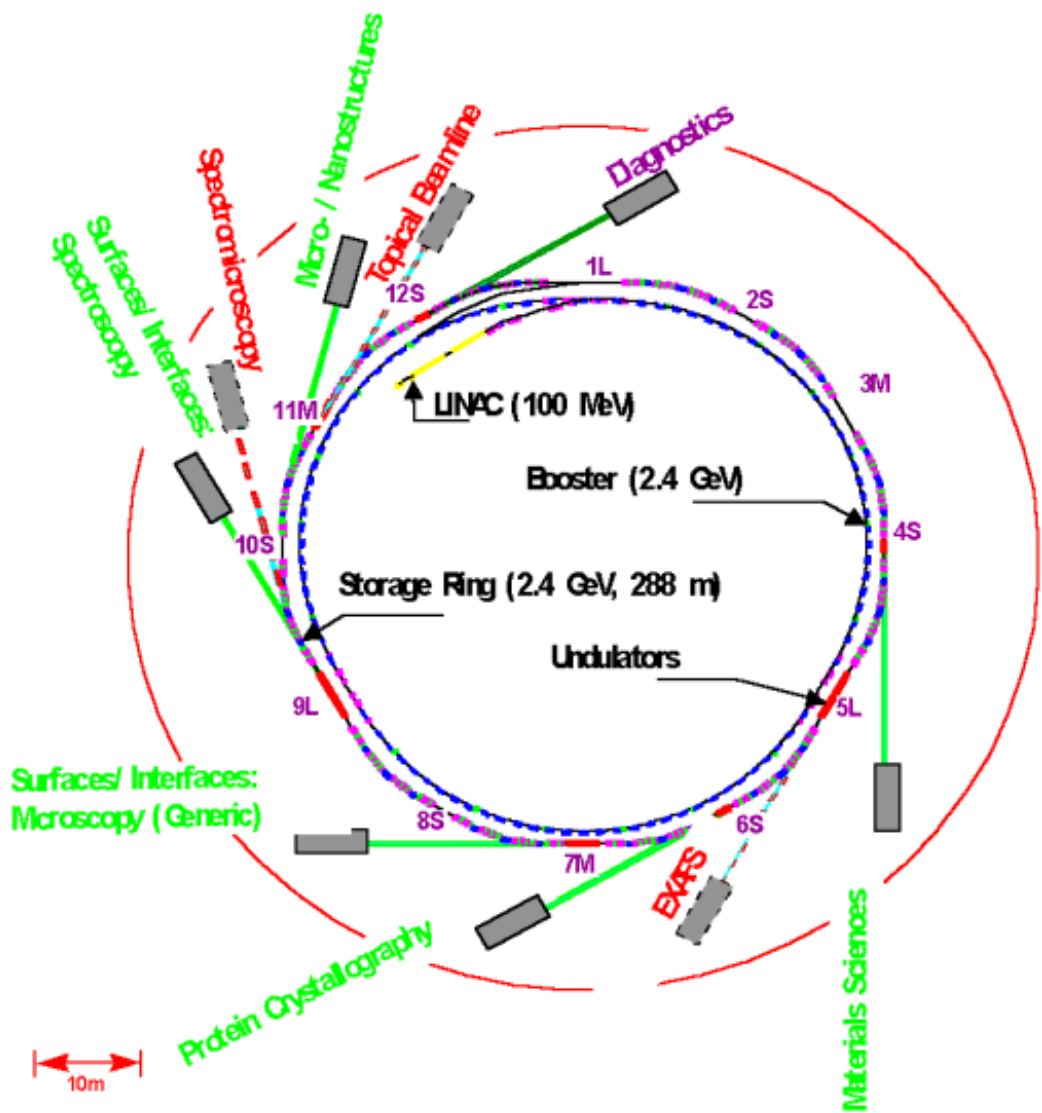

Figure 1: Layout of the SLS facility with pre-accelertor (LINAC), main accelerator (Booster), storage ring, and beamlines. In the first phase there are two beamlines for hard X-rays (Protein Crystallography and Material Science) and two for the soft VUV region (Surfaces/Interfaces Spectroscopy and Surfaces/Interfaces Microscopy).

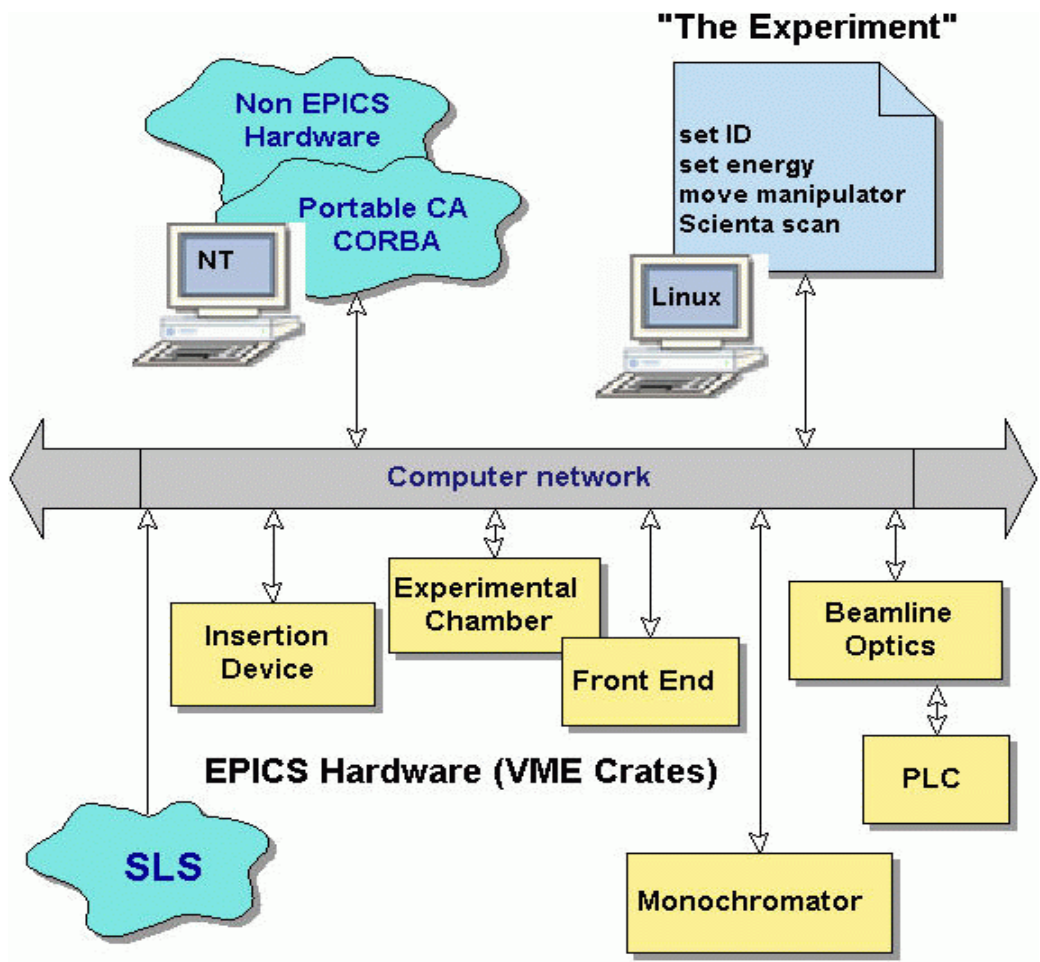

Figure 2: Schematical layout of a beamline experiment. The position and functionality of the VME crates follow the beamline topology (front-end, monochomator, beamline optics, experimental station 1, 2...) and define a beamline sub-system. The functionality of these sub-systems reside in VME crates seen on the lower part of the figure. On the upper part you can see the beamline clients doing the experiments. Since most of the functionality reside in VME crates (or virtual IOCs implemented with Portable Channel Access), an experiment may be implemented as a simple script.

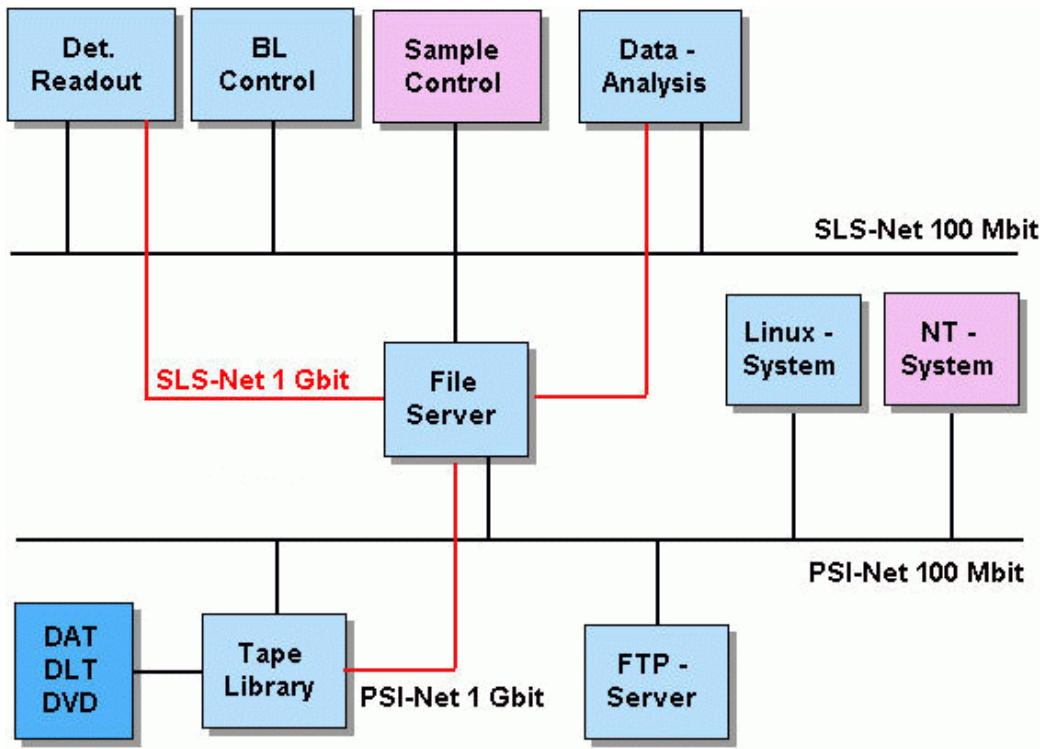

Figure 3: Schematical layout of data storage and handling.

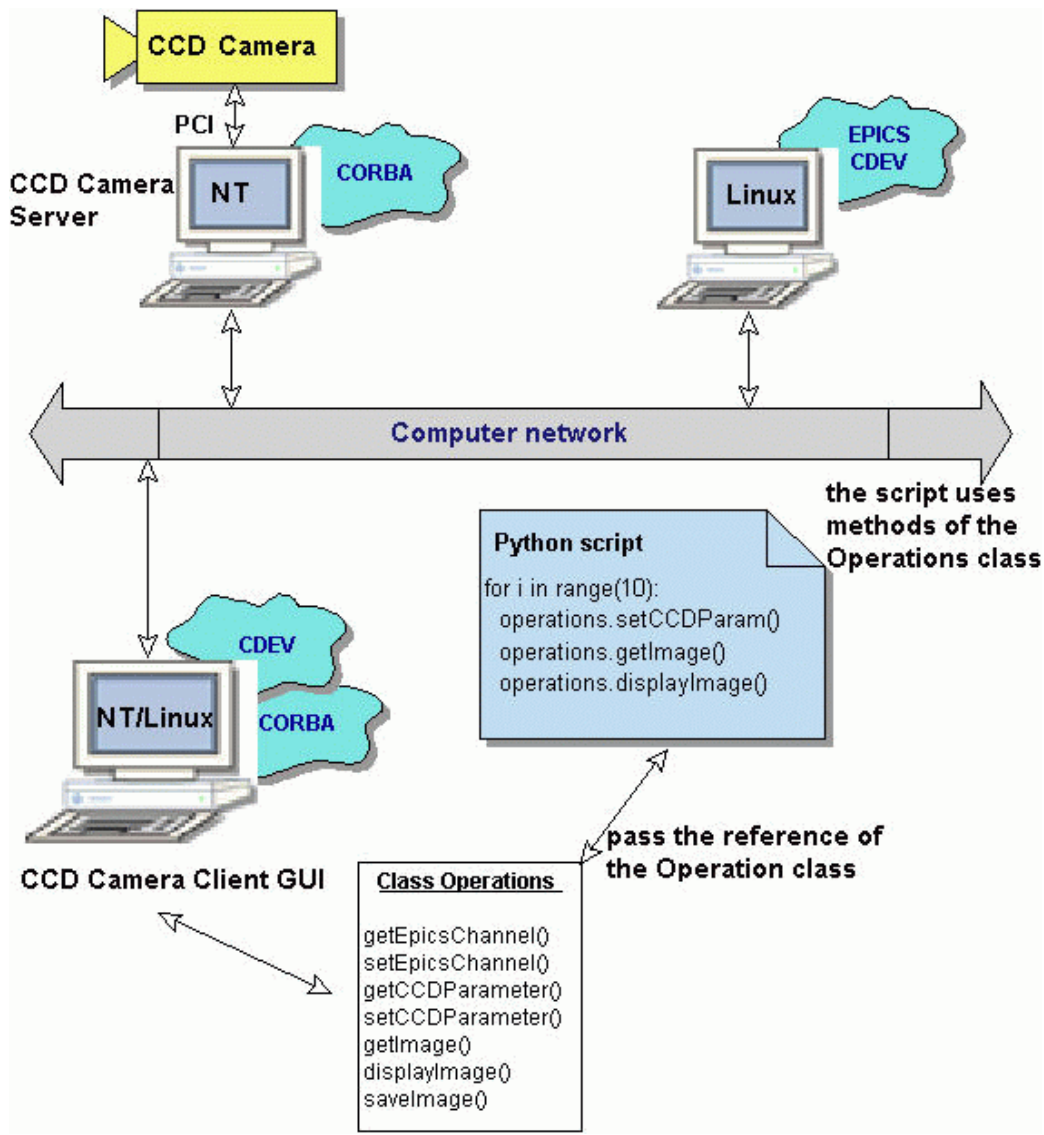

Figure 4: Schematical layout of a distributed architecture for controlling CCD camera including data acquisition (upper part) and treatment (lower part). The bulk of operations on CCD camera and EPICS control system is implemented in a class called "Operations". By interpreting the script (here written in Python) the reference of this class is passed to the script, which can call the class methods (containing image acquisition, parameter settings and accessing the EPICS environment).